\documentclass{article}
\usepackage{url}
\usepackage{longtable}
\usepackage{float}
\usepackage{subfloat}
\usepackage{lscape,epsfig}
\usepackage{graphicx}
\usepackage{amssymb}
\usepackage{amsmath}
\usepackage{natbib}
\textheight=22cm \DeclareSymbolFont{ppa}{OT1}{ppl}{m}{it}
\DeclareMathSymbol{\vv}{\mathalpha}{ppa}{'166}

minus 2.3mu \thickmuskip = 2.6mu plus 2mu minus 2.6mu

\let\svthefootnote\thefootnote

\begin{document}

\newcommand{\dd}{\,{\rm d}}
\newcommand{\ie}{{\it i.e.},\,}
\newcommand{\etal}{{\it $et$ $al$.\ }}
\newcommand{\eg}{{\it e.g.},\,}
\newcommand{\cf}{{\it cf.\ }}
\newcommand{\vs}{{\it vs.\ }}
\newcommand{\zdot}{\makebox[0pt][l]{.}}
\newcommand{\up}[1]{\ifmmode^{\rm #1}\else$^{\rm #1}$\fi}
\newcommand{\dn}[1]{\ifmmode_{\rm #1}\else$_{\rm #1}$\fi}
\newcommand{\upd}{\up{d}}
\newcommand{\uph}{\up{h}}
\newcommand{\upm}{\up{m}}
\newcommand{\ups}{\up{s}}
\newcommand{\arcd}{\ifmmode^{\circ}\else$^{\circ}$\fi}
\newcommand{\arcm}{\ifmmode{'}\else$'$\fi}
\newcommand{\arcs}{\ifmmode{''}\else$''$\fi}
\newcommand{\MS}{{\rm M}\ifmmode_{\odot}\else$_{\odot}$\fi}
\newcommand{\RS}{{\rm R}\ifmmode_{\odot}\else$_{\odot}$\fi}
\newcommand{\LS}{{\rm L}\ifmmode_{\odot}\else$_{\odot}$\fi}

\newcommand{\Abstract}[2]{{\footnotesize\begin{center}ABSTRACT\end{center}
\vspace{1mm}\par#1\par \noindent {~}{\it #2}}}

\newcommand{\TabCap}[2]{\begin{center}\parbox[t]{#1}{\begin{center}
  \small {\spaceskip 2pt plus 1pt minus 1pt T a b l e}
  \refstepcounter{table}\thetable \\[2mm]
  \footnotesize #2 \end{center}}\end{center}}

\newcommand{\TableSep}[2]{\begin{table}[p]\vspace{#1}
\TabCap{#2}\end{table}}

\newcommand{\FigCap}[1]{\footnotesize\par\noindent Fig.\  %
  \refstepcounter{figure}\thefigure. #1\par}

\newcommand{\TableFont}{\footnotesize}
\newcommand{\TableFontIt}{\ttit}
\newcommand{\SetTableFont}[1]{\renewcommand{\TableFont}{#1}}
\newcommand{\MakeTable}[4]{\begin{table}[htb]\TabCap{#2}{#3}
  \begin{center} \TableFont \begin{tabular}{#1} #4
  \end{tabular}\end{center}\end{table}}

\newcommand{\MakeTableSep}[4]{\begin{table}[p]\TabCap{#2}{#3}
  \begin{center} \TableFont \begin{tabular}{#1} #4
  \end{tabular}\end{center}\end{table}}

\newenvironment{references}%
{ \footnotesize \frenchspacing
\renewcommand{\thesection}{}
\renewcommand{\in}{{\rm in }}
\renewcommand{\AA}{Astron.\ Astrophys.}
\newcommand{\AAS}{Astron.~Astrophys.~Suppl.~Ser.}
\newcommand{\ApJ}{Astrophys.\ J.}
\newcommand{\ApJS}{Astrophys.\ J.~Suppl.~Ser.}
\newcommand{\ApJL}{Astrophys.\ J.~Letters}
\newcommand{\AJ}{Astron.\ J.}
\newcommand{\IBVS}{IBVS}
\newcommand{\PASP}{P.A.S.P.}
\newcommand{\Acta}{Acta Astron.}
\newcommand{\MNRAS}{MNRAS}
\renewcommand{\and}{{\rm and }}
\section{{\rm REFERENCES}}
\sloppy \hyphenpenalty10000
\begin{list}{}{\leftmargin1cm\listparindent-1cm
\itemindent\listparindent\parsep0pt\itemsep0pt}}%
{\end{list}\vspace{2mm}}

\def\TYLDA{~}
\newlength{\DW}
\settowidth{\DW}{0}
\newcommand{\dw}{\hspace{\DW}}

\newcommand{\refitem}[5]{\item[]{#1} #2%
\def\REFARG{#3}\ifx\REFARG\TYLDA\else, {\it#3}\fi
\def\REFARG{#4}\ifx\REFARG\TYLDA\else, {\bf#4}\fi
\def\REFARG{#5}\ifx\REFARG\TYLDA\else, {#5}\fi.}

\newcommand{\Section}[1]{\section{#1}}
\newcommand{\Subsection}[1]{\subsection{#1}}
\newcommand{\Acknow}[1]{\par\vspace{5mm}{\bf Acknowledgements.} #1}
\pagestyle{myheadings}

\newfont{\bb}{ptmbi8t at 12pt}
\newcommand{\xrule}{\rule{0pt}{2.5ex}}
\newcommand{\xxrule}{\rule[-1.8ex]{0pt}{4.5ex}}

\begin{center}
{\Large\bf
 The Cluster AgeS Experiment (CASE).\dag  \\
 Variable stars in the field of  
 the globular cluster M22}{\LARGE$^\ast$}
 \vskip1cm
  {\large
      ~~M.~~R~o~z~y~c~z~k~a$^1$,
      ~~I.~B.~~T~h~o~m~p~s~o~n$^2$,
      ~~W.~~P~y~c~h$^1$,
      ~~W.~~N~a~r~l~o~c~h$^1$,
              ~~R.~~P~o~l~e~s~k~i$^3$
      ~~and~~A.~~S~c~h~w~a~r~z~e~n~b~e~r~g~~--~~C~z~e~r~n~y$^1$
   }
  \vskip3mm
{ $^1$Nicolaus Copernicus Astronomical Center, ul. Bartycka 18, 00--716 Warsaw, Poland\\
     e--mail: (mnr, pych, wnarloch, alex)@camk.edu.pl\\
  $^2$The Observatories of the Carnegie Institution for Science, 813 Santa Barbara
      Street, Pasadena, CA 91101, USA\\
     e--mail: ian@obs.carnegiescience.edu\\
  $^3$ Department of Astronomy, Ohio State University, 140W. 18th Ave., Columbus, OH43210, USA\\
     e--mail: rpoleski@astrouw.edu.pl}
\end{center}

\vspace*{7pt}
\Abstract 
{The field of the globular cluster M22 (NGC 6656) was monitored between 2000 and 2008 in a search 
for variable stars. $BV$ light curves were obtained for 359 periodic, likely periodic,
and long--term variables, 238 of which are new detections. Thirty nine newly detected variables, 
and 63 previously known ones are members or likely members of the cluster, including 
20 SX~Phe, 10 RRab and 16 RRc-type pulsators, one BL Her-type pulsator, 21 contact binaries, and 9 detached 
or semi--detached eclipsing binaries. The most interesting among the identified objects are V112 --
a bright multimode SX Phe pulsator,  V125 -- 
a $\beta$ Lyr--type binary on the blue horizontal branch, V129 -- a blue/yellow straggler with a 
W~UMa--like light curve, located halfway between the extreme horizontal branch and red giant branch, 
and V134 -- an extreme horizontal branch object with $P=2.33$ d and a nearly sinusoidal light curve; all 
four of them are proper motion (PM) members of the cluster. Among nonmembers, a $P=2.83$ d detached eclipsing 
binary hosting a $\delta$ Sct-type pulsator was found, and a peculiar $P=0.93$ d binary with ellipsoidal 
modulation and narrow minimum in the middle of one of the descending shoulders of the sinusoid. We 
also collected substantial new data for previously known variables; in particular we revise the statistics 
of the occurrence of the Blazhko effect in RR Lyr-type variables of M22.  
}
{globular clusters: individual (M22) -- stars: variables -- 
stars: SX Phe -- blue stragglers -- binaries: eclipsing
}

\let\thefootnote\relax\footnotetext{\dag CASE was initiated and for long time led
by our friend and tutor Janusz Kaluzny, who prematurely passed away in March 2015.}
\let\thefootnote\relax\footnotetext
{$^{\mathrm{\ast}}$Based on data obtained with the Swope telescope at  
Las Campanas Observatory.}
\let\thefootnote\svthefootnote

\Section{Introduction} 
\label{sec:intro}
M22 is projected against the Galactic bulge at $l=9\zdot\arcd9$, $b=-7\zdot\arcd6$, in 
a substantially reddened region with $E(B-V)$ varying between 0.26 mag and 0.39 mag 
across our field of view\footnote{The extinction calculator at 
http://ned.ipac.caltech.edu/help/extinction\_law\_calc.html was used}. Core radius $r_c$, half--light 
radius $r_h$, tidal radius $r_t$, [Fe/H] index, radial velocity, heliocentric distance 
$d_\odot$, and galactocentric distance $d_G$ of the cluster are equal to 1\zdot\arcm33, 3\zdot\arcm36, 
32\zdot\arcm0, -1.70, -146.3$\pm$0.2 km/s, 3.2 kpc and 4.9 kpc, respectively (Harris 1996, 
2010 edition). Dotter et al. (2010) excluded M22 from their age survey "because it is 
known to harbor multiple stellar populations". Indeed, Lee (2016) suggests that it is 
a merger of two globular clusters (GCs) which occurred in a dwarf galaxy, subsequently 
accreted onto the Milky Way. This might explain differences in age estimations of M22, 
varying from 12--13 Gyr (Lee 2015, 2016) up to 14 Gyr (Marino et al. 2009).  

M22 is classified as an old GC of Oosterhoff type II, and has a rich and long blue 
horizontal branch (BHB). The $(N_{BHB}-N_{RHB})/(N_{BHB}+N_{RHB}+N_{RR})$ index, where 
$N_{BHB}$ is the number of BHB stars, $N_{RHB}$ the number of red HB stars (located 
redward of the instability strip on the CMD), and $N_{RR}$ the number of RR Lyr stars, 
is equal to $0.97\pm0.1$,  one of the largest among GCs with a substantial 
population of RR Lyr pulsators (Kunder et al. 2013a, hereafter K13). 

Even though M22 is one of the closest GCs to the Sun, factors like very strong 
contamination of its field by bulge stars, substantial differential extinction, 
and appreciable concentration ($c=\log r_t/r_c = 1.38$; Harris 1996, 2010 edition) 
make it a rather challenging target for studies. The pre--CCD searches for variables, 
summarized by Clement et al. (2001; 2017 edition\footnote{\url{
http://www.astro.utoronto.ca/~cclement/cat/C0100m711}}) (hereafter C01-17), 
resulted in the detection of 43 objects. The targeted CCD surveys performed so 
far (Kaluzny \& Thompson 2001, hereafter KT01; Pietrukowicz \& Kaluzny 2003, hereafter 
PK03; K13, and Sahay, Lebzelter \& Wood 2014) brought additional 56 discoveries, 
including two optical cataclysmic variables 
and a microlensing event (Pietrukowicz et al. 2005, 2012). Fourteen of these objects 
are listed by C01-17 as members or possible members of the cluster, including eight 
RR Lyr pulsators, one contact binary, and five semiregular variables. 

Apart from 
normal stars, the cluster contains two millisecond pulsars (Lynch et al. 2011), and   
two candidate stellar--mass black holes (Strader et al. 2012). Finally, 
according to Kains et al. (2016) M22 provides the best chance to detect an 
intermediate--mass black hole via astrometric microlensing.

Our survey is a part of the CASE project (Kaluzny et al. 2005) conducted using 
telescopes of the Las Campanas Observatory, with an aim of increasing the inventory 
of variable objects in the field of M22. It completes the previous findings of KT01
(based on 76 frames obtained during one night on the du Pont telescope) and PK03
(based on 31 archival HST/WFPC2 frames, and necessarily limited to the central part
of the cluster). Altogether we identified 283 periodic, likely periodic or long--term 
variables not cataloged by C01-17, of which 45 were independently found by Soszy\'nski 
et al. (2016; hereafter S16) during 
the OGLE--IV survey of the Galactic bulge. In Section~2, we briefly report on the 
observations and explain the methods used to calibrate the photometry. Newly discovered 
variables are presented and discussed in Section~3. Section~4 contains new data on 
previously known variables which we consider worthy of publishing, and the paper is 
summarized in Section~5.
\section{Observations and data processing}
\label{sec:obs}
Our paper is based on images acquired with the 1.0--m Swope telescope equipped with the 
$2048\times 3150$ SITe3 camera. The field of view was $14.8\times 22.8$ arcmin$^2$ 
at a scale of 0.435 arcsec/pixel. Observations were conducted on 86 nights from 
April 11, 2000 to August 22, 2008; always with the same set of filters.
A total of 2730 $V$--band images and 384 $B$--band images were selected for the
analysis. The seeing ranged from $1''.2$  to $3''.6$ and $1''.2$  to $3''.9$  for 
$V$ and $B$, respectively, with median values of $1''.4$  in both filters.

The photometry was performed using an image subtraction technique implemented in the 
DIAPL package.\footnote{Available from http://users.camk.edu.pl/pych/DIAPL/index.html} 
To reduce the effects of PSF variability, each frame was divided into 4$\times$6  
overlapping subframes. The reference frames were constructed by combining 18 images 
in $V$ and 17 in $B$ with an average seeing of 1.$''$1 and 1.$''$2, respectively.
The light curves derived with DIAPL were converted from differential counts to magnitudes 
based on profile photometry and aperture corrections determined separately for each 
subframe of the reference frames. To extract the 
profile photometry from reference images and to derive aperture corrections, the 
standard Daophot, Allstar and Daogrow (Stetson 1987, 1990) packages were used. 
Profile photometry was also extracted for each individual image, enabling useful 
photometric measurements of stars which were overexposed on the reference frames. 
\subsection{Calibration}

The photometric calibration is based on standard magnitudes and colors derived by KT01. 
Using over 40,000 comparison stars common to our survey and theirs, the following 
transformation to the standard system was derived:
\begin{align}
  V &= v + 2.0763(2) + 0.0191(2)\times(b-v) \nonumber\\
  B - V &= -0.1504(3) + 1.0400(3)\times(b-v) \nonumber,
\end{align}
where lower case and capital letters denote instrumental and standard magnitudes, 
respectively, and numbers in parentheses are uncertainties of last significant 
digits.

\begin{figure}[H]
   \centerline{\includegraphics[width=0.95\textwidth]{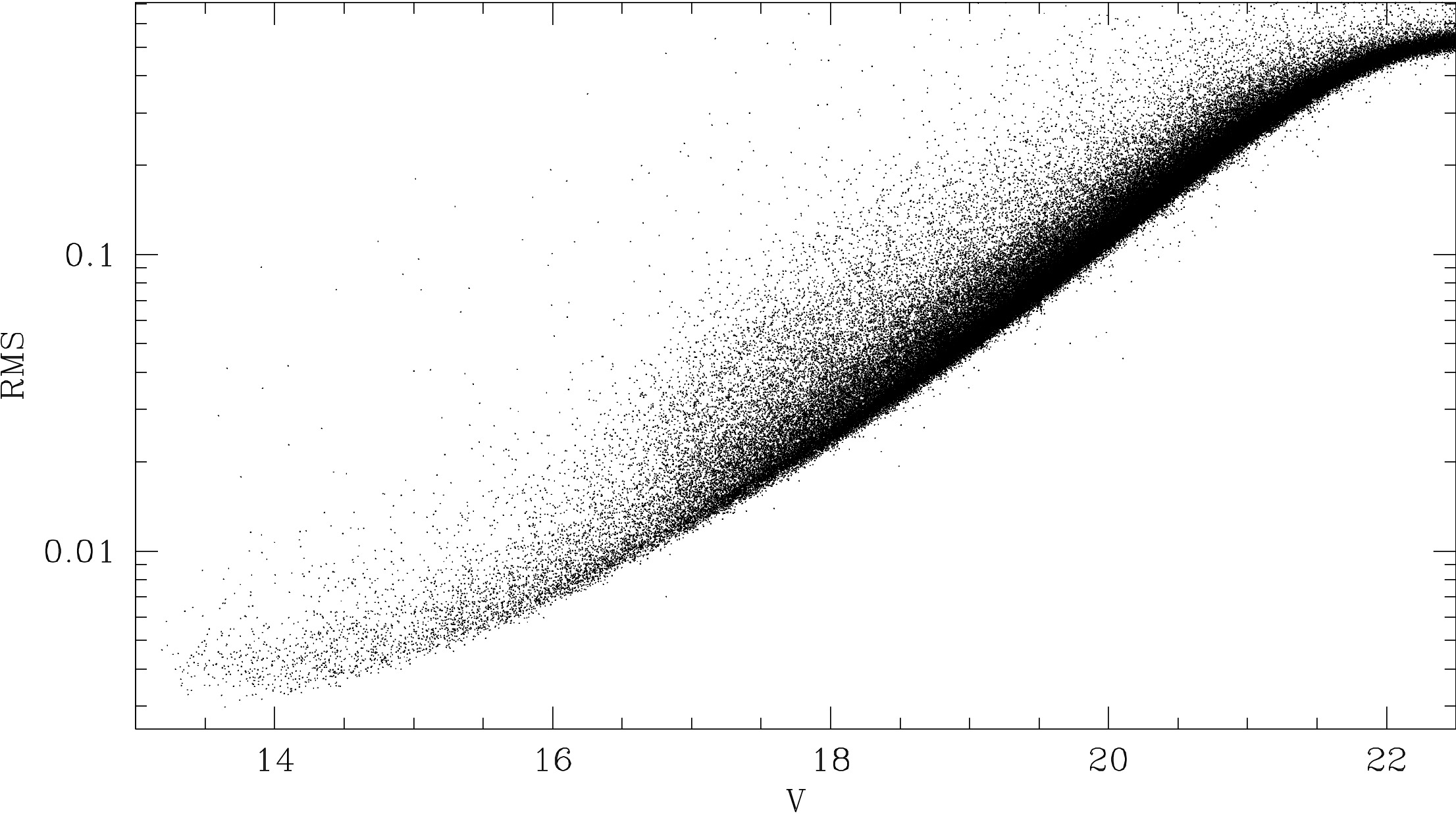}}
   \caption{ Standard deviation vs. average $V$-band magnitude for
    light curves of stars from the M22 field. 
    \label{fig:rms}}
\end{figure}

\begin{figure}[H]
   \centerline{\includegraphics[width=0.95\textwidth]{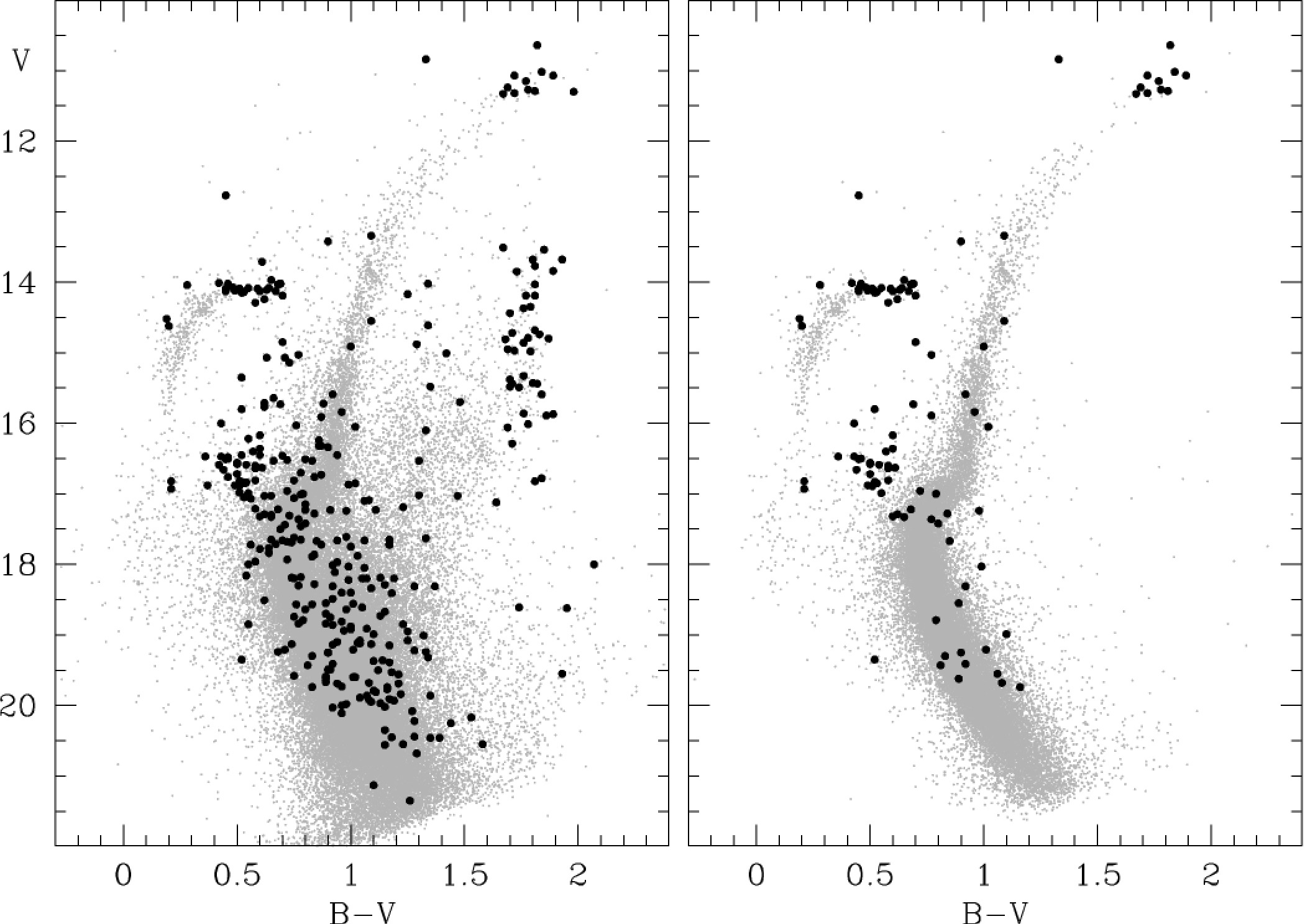}}
   \caption{CMD for the observed field. Left: all stars 
    for which proper motions were measured. Black points mark all the variables detected 
    within the present survey for which $B$-band magnitudes were available. Right: same 
    as in the left panel, but for PM-members of the cluster only. 
    \label{fig:cmds}}
\end{figure}

Crowding in the field of view resulted in enhanced blending, which in turn 
significantly increased the scatter of photometric measurements in the observed 
magnitude range. For example, observations made of the globular cluster NGC~3201
(using the same instrument setup) resulted in a smallest scatter of 0.1 mag at
$V$ = 21 mag (Kaluzny et al. 2016), while
for M22 the best photometric accuracy is  $\sim$0.23 mag (Fig. \ref{fig:rms})
at the same brightness level. Fig.~\ref{fig:cmds}, 
shows the CMD of the observed field and was constructed based on the reference images. 
To make the figure readable, only stars with measured proper motions (Narloch et al. 2017;
hereafter N17) are selected to serve as a background for the variables. Stars 
identified as proper--motion (PM) members of the cluster are shown in the right panel.
 
\subsection{Search for variables}

The search for periodic variables was conducted using the AOV and AOVTRANS algorithms 
implemented in the TATRY code (Schwarzenberg--Czerny 1996 and 2012;  
Schwar\-zenberg--Czerny \& Beaulieu 2006). We examined time--series photometric data 
of 132,457 stars brighter than $V\sim$22 mag. The photometric 
accuracy was partly offset by the large number of available frames, and as a 
result we were able to detect periodic signals with amplitudes of $\sim$0.02 mag 
down to $V\approx15$ mag, and $\sim$0.1 mag down to $V\approx21$ mag.

Among the known variables within our field of view, light curves were obtained for 
all 45 stars discovered by S16, and for 76 out of 85 C01-17 stars. Of the latter,
light curves are missing for SLW-7 which was overexposed in our frames, and for seven 
PK03 stars located close to the center of the cluster. We identified 238 new variable 
or likely variable stars, 36 of which are PM--members or likely PM--members of M22. 
Membership status was also assigned to the variables known before.
\footnote{Data for all the identified variables are available
at http://case.camk.edu.pl} 

\section{The new variables}

\begin{figure}
   \centerline{\includegraphics[width=0.95\textwidth]{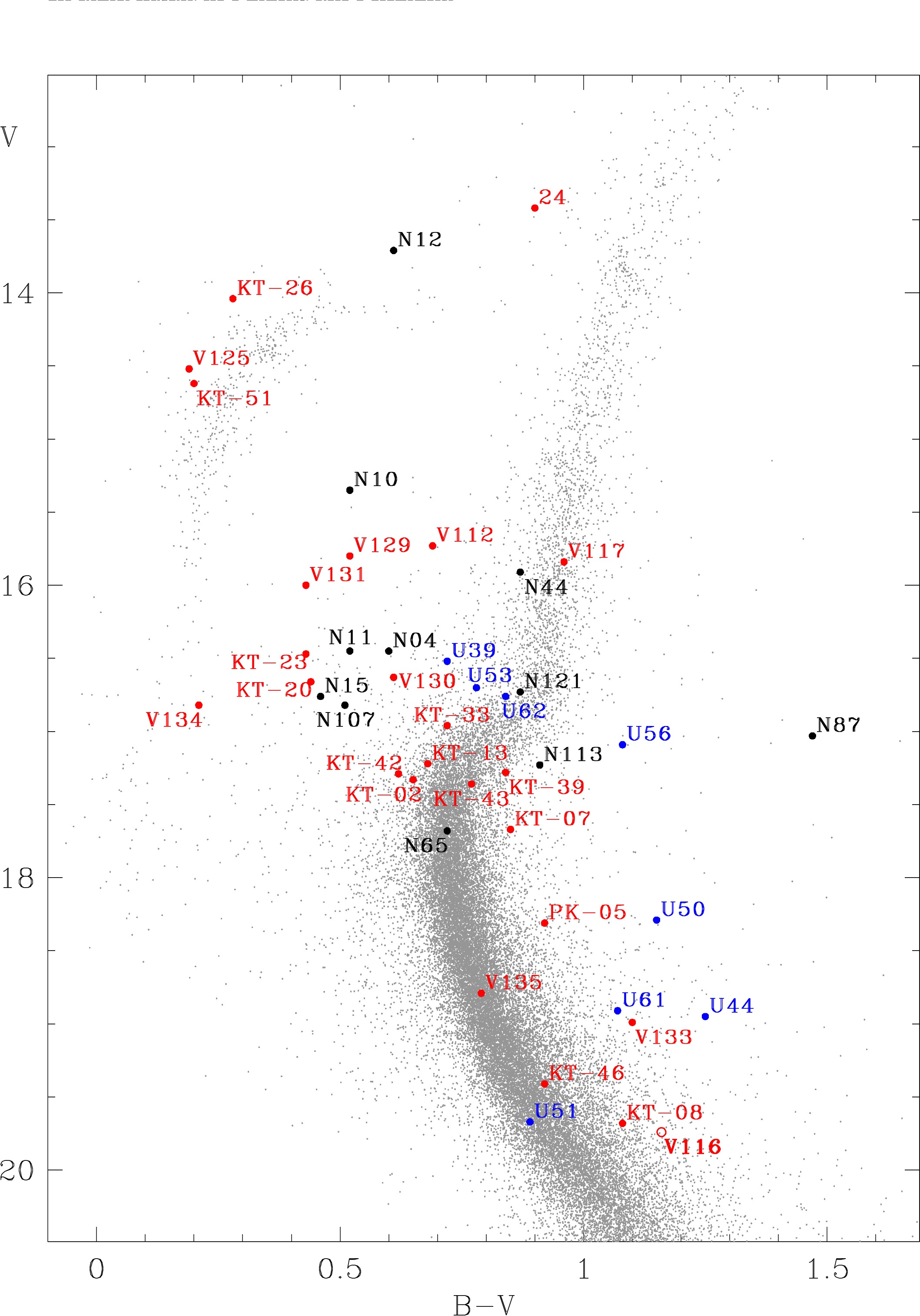}}
   \caption{CMD for the observed field with locations of the variables described  
    in Sections 3 and 4 (to make the figure readable, the remaining variables 
    identified within the present survey are 
    not shown). Red: PM-members of M22; blue: stars for which PM-data
    are missing; black: field stars. Filled circles: confirmed variables;
    open circle: suspected variable. The gray background stars are the 
    same as in the right panel of Fig.~\ref{fig:cmds}.
    \label{fig:cmd_var}}
\end{figure}

Basic data for selected variables not listed in C01-17 are given in 
Table \ref{tab:CASE}.  For our naming convention to agree with 
that of C01-17 we start numbering the new variable cluster members from V102. The 
remaining variables are given names from U01 on (stars for which no PM--data are present)
and from N01 on (stars whose PM indicates that they do not belong to M22). The equatorial 
coordinates in columns 2 and 3 conform to the UCAC4 system (Zacharias et al. 2013),
and are accurate to 0$''$.2 -- 0$''$.3. The $V$--band magnitudes in column 4 correspond 
to the maximum light in the case of eclipsing binaries; in the remaining cases 
the average magnitude is given. Columns 5--7 give $B-V$ color, amplitude in the 
$V$--band, and period of variability. A CMD of M22 with locations of 
the variables is shown in Fig.~\ref{fig:cmd_var}. Field objects are marked in black, 
those for which the PM data are missing or ambiguous in blue, and members of the cluster 
in red. The gray background stars are the PM--members of M22 from the right panel of 
Fig.~\ref{fig:cmds}. 

\subsection {Members and likely members of M22}
\label{sec:newmem}

\begin{figure}
   \centerline{\includegraphics[width=0.95\textwidth]{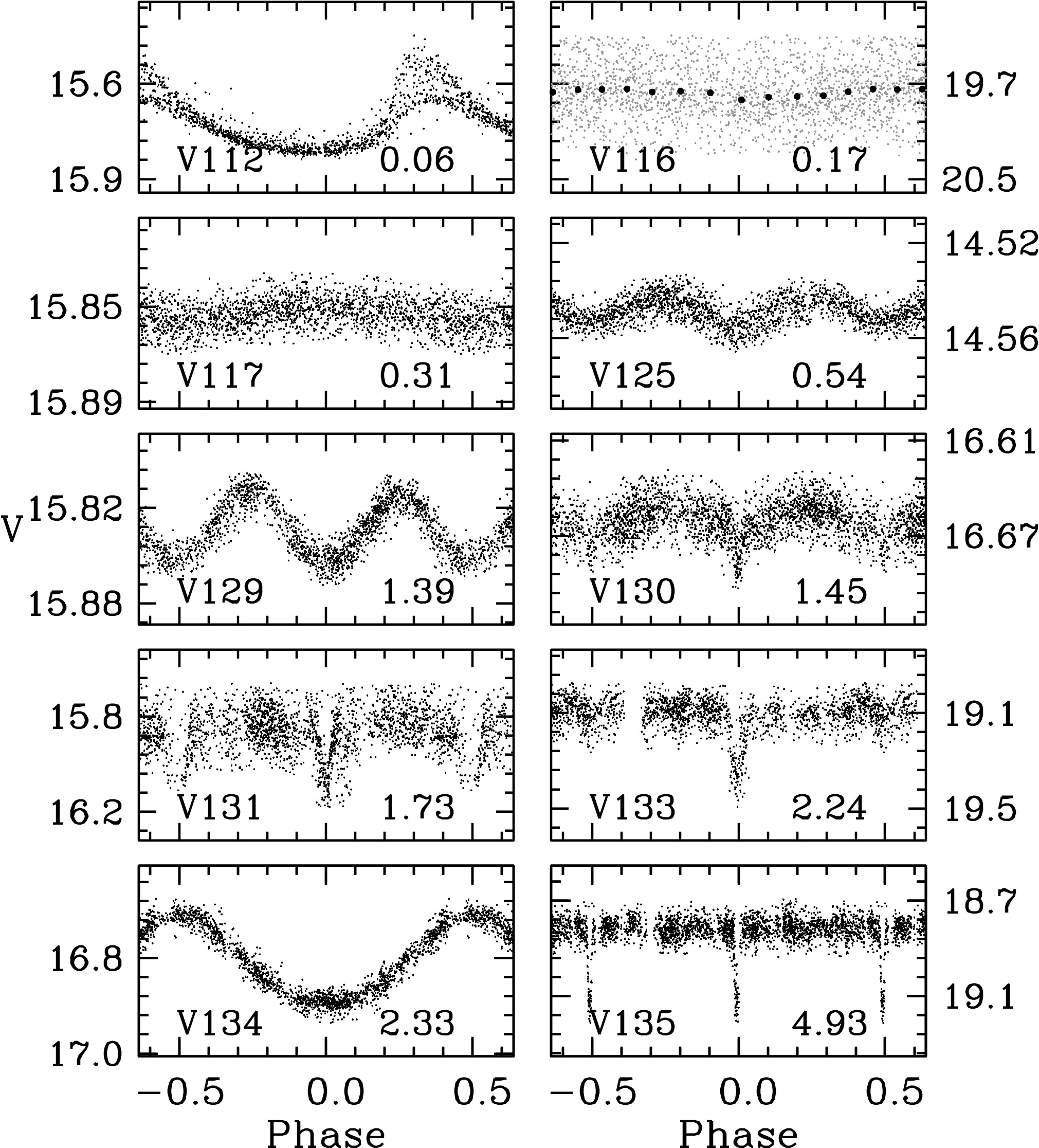}}
   \caption{Phased $V$-band light curves for a selection of the newly detected 
    variable members or likely members of M22. Phase-binned data are shown for 
    V116 with heavy black points. Individual panel labels give star ID, and period 
    in days.
    \label{fig:CASE_Yfin}}
\end{figure}

Based on proper motions, distances from the center of the cluster, and CMD locations
we identified 39 M22--members not cataloged by C01-17 (among them, three discovered 
by S16). A star was considered a member or likely member if
one of the following criteria was fulfilled: 
\begin{enumerate}
\item PM--membership probability $P_{PM}\geq$70\%.
\item $P_{PM}<$70\%, but CMD-location compatible with cluster membership, variability 
      type compatible with CMD--location, and geometric 
      membership probability $P_{geom}=1-\pi r^2/S>90$\%, where $r$ is star's distance 
      from the center of M22 ($\alpha$ = 18$^{\mathrm h}$ 36$^{\mathrm m}$ 
      23\zdot$^{\mathrm s}$94, $\delta$ = -23\arcd 54\arcm 17\zdot\arcs1) in arcseconds, and 
      $S=1.22\times10^6$ is the size of the 
      field of view in arcseconds$^2$ (there are two such cases).
\item Proper motion not known, but $P_{geom}>70$\%, CMD-location compatible with cluster 
      membership, and variability type compatible with CMD--location.
\end{enumerate}
Details concerning PM measurements and calculations of membership probability are given 
in N17, who also provide a PM catalog for nearly 450000 stars in the fields of 12 GCs.
In the following, we describe the ten most interesting variables, whose light curves 
are shown in Fig. \ref{fig:CASE_Yfin}. 

Our data suggest that multimode pulsations are likely in 16 SX Phe stars (seven new ones
and nine from the C01-17 catalog). The most interesting one among them is the newly detected
variable V112, which is also the brightest and reddest blue straggler (BS). It clearly 
exhibits multimode pulsations at an amplitude of $\sim$0.3 mag suitable for asteroseismology
analysis which in turn would provide valuable information on BS mass. Admittedly,
its CMD location may seem a bit extreme for this type of variability, however both the $V$
and $B$ lightcurves are of good quality, so that a large error in $<B>-<V>$ can be excluded. 
$P_{PM}=$100\% for V112, however we feel a radial velocity measurement would be necessary 
to confirm its membership. The star is a component of a blend. However, in the archival HST 
frame NGC6656-J9L948010 V112 is much brighter than the remaining components (in fact, it is
strongly overexposed). 

V116, a sinusoidal variable on the lower main sequence, is a 100\% PM--member of M22. Our light 
curve is of poor quality because of partial blending with a much brighter star $\sim1''.5$ 
distant. We did not detect any periodicity in the latter, and V116 is well isolated in the 
archive HST frame NGC6656-U2X80302T. Thus, if the weak periodic signal we observe is real, 
then it must originate in V116 (not being entirely sure about its reality, we marked the star 
as a suspected variable). V116 would then closely resemble the optical counterpart of the 
X--ray source CX1 in M4 (Kaluzny et al. 2012). 

V117 is a low--amplitude sinusoidal variable with a short period (0.31 d) clearly incompatible 
with its location on the red giant branch (RGB). However, it is a 100\% PM--member of M22. Our image 
of V117 is perfectly symmetric, but, since the star is located in the unobserved by HST part 
of the cluster, the possibility of blending cannot be excluded. If adaptive optics photometry
confirmed that we deal with a single light source, V117 would become an interesting target for
further research. 

V125, located on the blue horizontal branch (BHB), has a $\beta$~Lyr--type (EB) light curve with minima 
of different depths, and $P_{PM}=$100\%. No HST data are available for this object. The star 
is well separated from its neighbors; nevertheless adaptive optics would be needed to exclude 
blending. If not a blend, V125 would be one of the very rare BHB binaries with short periods 
(Heber 2016). 

V129, a BS with $P_{PM}=$100\%, which exhibits a W UMa--like (EW) light curve with minima of 
different depth, is peculiar because of its long period (1.39 d). The observed minima are broader 
than the maxima, also not fitting a W UMa interpretation. In our frame, variable  V129 is blended 
with at least two fainter stars; unfortunately their contribution to the total light cannot be 
estimated because of lacking HST data.  

The blue stragglers V130 and V131 are Algol-type eclipsing (EA) binaries with a strong ellipsolidal
effect. No large 
observational effort would be needed to obtain reasonable quality light and velocity curves 
for these systems, and determine their parameters. Such a project would be worthwhile, as BS 
Algols provide a very demanding test suite for stellar evolution codes even in cases when 
their parameters are not accurately known (St\c{e}pie\'n, Pamyatnykh \& Rozyczka 2017). 

V133 is a detached eclipsing binary with a period ambiguity. $P_1=2.244228$~d in Table~\ref{tab:CASE}
is the best fit to the light curve, with only one minimum visible. For $P_2=1.195288$ d a secondary
minimum appears, which may be as deep as the primary minimum. However, the fit becomes markedly poorer.
Since PM is not available for V133, and $P_{geom}=90$\%, V133 is just a likely member of M22; potentially
interesting since it might serve as age and distance indicator if its membership were confirmed. If
it belongs to M22, the absence of the second minimum speaks against $P_1$, as the system is located
too high above the lower main sequence for such a large luminosity difference between the components.

V134 is a nearly sinusoidal variable discovered by S16 (their star OGLE-BLG-ECL-423136). With 
$P_{PM}=$100\%, $P=2.33$ d, and a location between the extreme horizontal branch (EHB) and the BS region
on the CMD, it constitutes a real puzzle. The high quality  $B$ and $V$ light curves yield a reliable $<B>-<V>$,
so that the chance that V134 is horizontally misplaced in the CMD is low. The most natural cause 
of this type of variability is a strong reflection effect similar to that observed in HW Vir binaries, 
however the period of V134 (2.331 d) is much longer than the longest period known among the members
of that class ($\sim$0.75 d; Heber 2016). A slight 
elongation of the image of this star in our frames suggests a tight blend;
unfortunately no HST data are available. Clearly, a spectroscopic follow--up is needed to 
verify its membership and reveal its nature. 

V135, another detached eclipsing binary with  a 1:2 period ambiguity, is located on the lower main
sequence. $P=4.928$ d and $P=2.464$ d fit the lightcurve almost equally well; however the 
longer period implies nearly the same brightness of the components, which is barely 
compatible with the CMD location of the system. Thus, although V135 is a 100\% PM--member of M22, 
its membership should be verified through radial velocity measurements.

\subsection{Stars of unknown PM--membership}
\label{sec:nomemdata}

\begin{figure}
   \centerline{\includegraphics[width=0.95\textwidth]{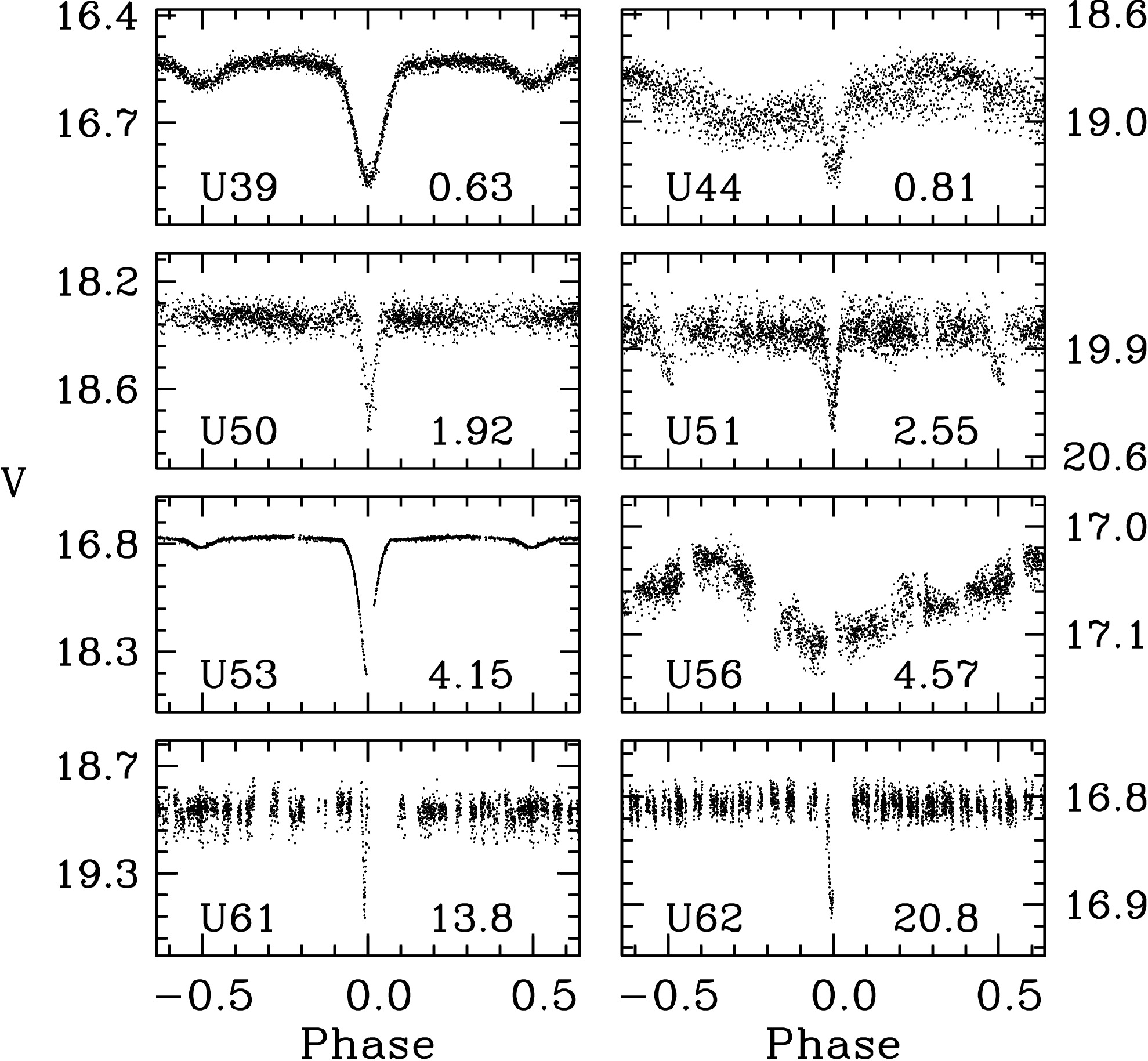}}
   \caption{Phased $V$-band light curves for a selection of the new variables 
    from the observed field for which no PM-data are present. Individual 
    panel labels give star ID, and period in days.
    \label{fig:CASE_Ufin}}
\end{figure}

In our sample, there are 69 variables with $P_{geom}<70$\% and unknown proper motions, at 
least some of which may turn out to belong to M22. Below we describe eight of the most interesting
cases, whose light curves are shown in Fig. \ref{fig:CASE_Ufin}.

Algols U39 (OGLE-BLG-ECL-423130) and U53 are prospective yellow stragglers. If their 
membership is confirmed they will provide excellent opportunity to test and/or calibrate 
stellar evolution codes (St\c{e}pie\'n, Pamyatnykh \& Rozyczka 2017).

U44 is a RS CVn--type eclipsing binary with a strong sinusoidal modulation, resembling
V9 in NGC 6971 (Kaluzny 2003; Bruntt et al. 2003) or a sample of RS CVn discovered within
the OGLE III survey and described by Pietrukowicz et al. (2013). Only one eclipse is 
visible, situated almost in the middle of the ascending branch of the light curve. The 
modulation originates from spot(s) possibly accompanied by mass transfer effects, 
similarly to those observed in R Ara (Baki\c{s} et al. 2016). Since such systems are 
rare, a follow--up of U44 would be desirable independently of its membership status.

U50 and U61 are detached eclipsing binaries located to the right of the lower main 
sequence. Both their light curves  reveal only one eclipse. If  follow--up 
photometry confirms our light curve fits, the systems would become interesting 
red straggler candidates (see e.g. Kaluzny 2003).  

U51, a detached eclipsing binary with two eclipses visible, has a period long enough
(2.6 d) to serve as age and distance indicator despite its low brightness.

U56, located redward of the subgiant branch, is another red straggler candidate.

U62 is a detached eclipsing binary located on the subgiant branch, and another 
potential excellent age and distance indicator. As our light curve covers only a part 
of a single eclipse, its period of 20.8d is only tenative.

\subsection{Field variables}
\label{sec:fieldvar}
\begin{figure}
   \centerline{\includegraphics[width=0.95\textwidth]{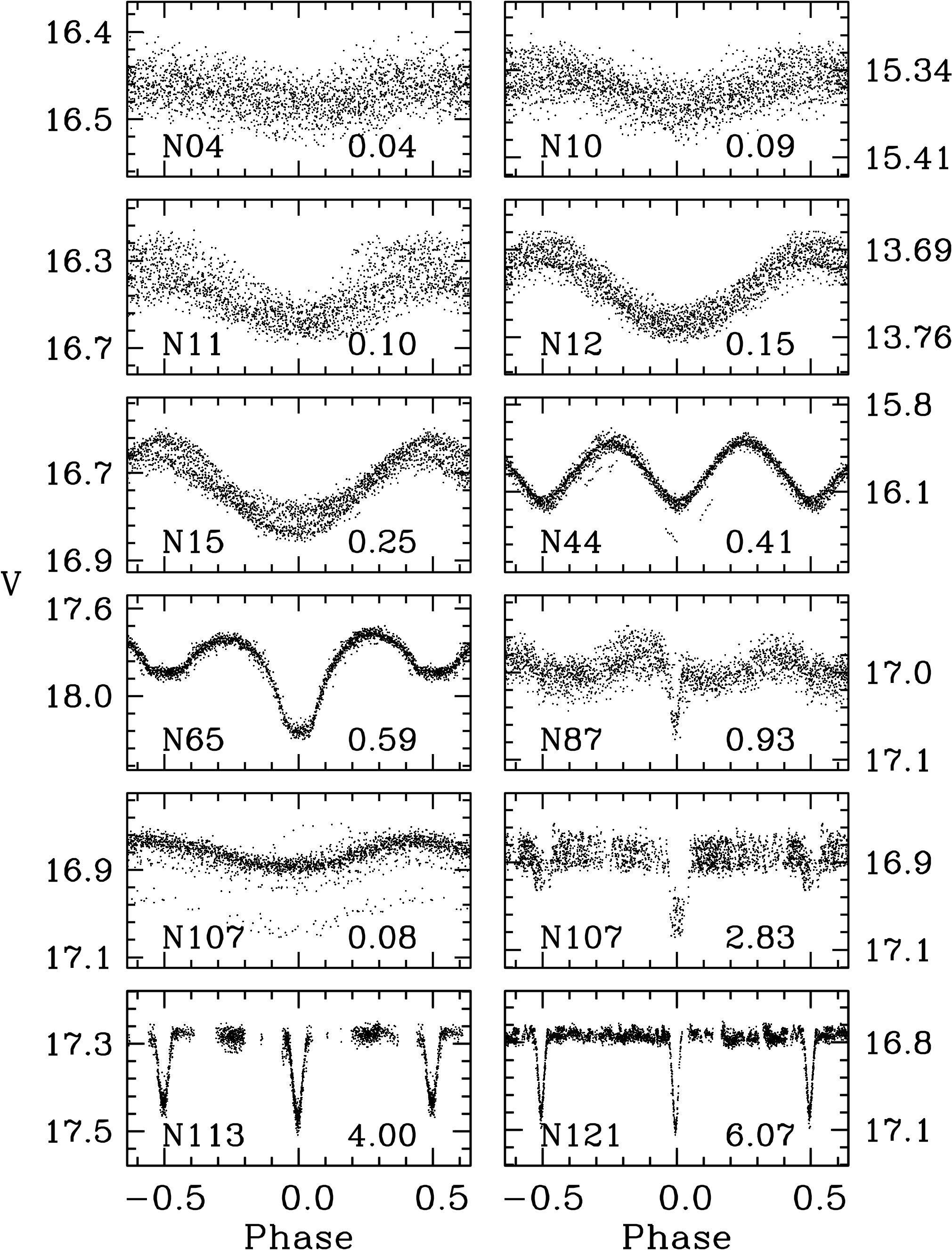}}
   \caption{Phased $V$-band light curves for a selection of the new variables
    from the observed field whose proper motions indicate that they do not 
    belong to M22. Individual panel labels give star ID and period in days.  
    N107 is phased with the pulsation period of its $\delta$ Sct / SX Phe 
    component, and with the orbital period.
    \label{fig:CASE_Nfin}}
\end{figure}

We identified 176 variables which according to N17 do not belong to M22. As errors 
in PM measurements cannot be entirely excluded, a few of them may in principle turn 
out to be cluster members. For that reason, while selecting the most interesting cases, 
we paid special attention to stars located on the CMD in the vicinity of the turnoff 
or in the BS region. The light curves of the selected variables are shown in 
Fig.~\ref{fig:CASE_Nfin}. 

N04, N10, and N11 are either field $\delta$ Sct variables or cluster SX Phe stars
and blue stragglers, all showing clear multimode pulsations. 

N12, a clear multimode pulsator located in the RR Lyr gap, has a period of only 0.15 d, 
which unambiguously identifies it as a field $\delta$ Sct star. 

N15, located in the BS region, is another multimode pulsator. Its period of 0.25 d is 
too long for a SX Phe variable; therefore it must also be a field $\delta$ Sct star. 

N44 is a field contact binary with a variable light curve. Its brightness seems to have 
decreased by $\sim$0.08 mag between 2000 and 2008 (the 2008 data were collected during 
four nights, so that a zero point artefact is rather unlikely -- the more that such 
effects are not seen in any other lightcurve).

N65 (OGLE-BLG-ECL-423254) is a W UMa eclipsing binary in poor thermal contact. 
The secondary eclipse is total, allowing 
an estimationi of the temperature of the primary from the color--temperature calibration. The 
observed $B-V$ index is 0.70 mag. Assuming a reddening of 0.30 mag (an average for M22) and using 
the calibration of Sousa et al (2011) one obtains $T_1=6500$ K. An approximate solution 
of the $V$- and $B$-band light curves with the PHOEBE implementation of the Wilson--Devinney code
(Pr\v{s}a and Zwitter 2005) yields $i=85\zdot\arcd8$, $T_2=4400$ K, and $\Delta M_{bol}=2.9$ 
mag between the components. Neglecting the contribution of the secondary, and assuming that 
N65 is a member of M22, from the observed $V = 17.85$ mag at maximum light, we obtain $M_{bol}^1
=4.38$ mag. This absolute brightness is reproduced by a W--D solution with semimajor axis and mass
of the primary of 2 $R_\odot$ and 0.25 $M_\odot$, respectively. Since the latter value is  
much too low for a 6500 K star, N65 must be a background object, only interesting because of
the significant temperature difference between the components. 

N87 seems similar to  U44, however there is a significant difference between them: N88 has 
two maxima per period instead of one. Since a configuration of two nearly identical 
spots at locations differing by nearly 180$^\circ$ in longitude is rather unlikely, the
nature of N88 is puzzling; the more that a similar object, OGLE-GD-ECL-04649, mentioned by 
Pietrukowicz et al. (2013), exhibits both a single and a double maximum at various seasons. 
The double-peaked curve resembles that of a cataclysmic variable with a giant donor (e.g. 
T CrB) yet the color is 1 mag too red. The system clearly deserves thorough follow--up 
observations, especially since  it is just 1.4 arcsec distant from the Chandra 
X--ray source C183656.05-234845.5 with $(\alpha,\delta)_{2000}$ = (279.23355, -23.81263).

N107 is a detached eclipsing binary, interesting independently of its membership status, since 
it hosts a $\delta$ Sct or SX Phe star. In Fig.~\ref{fig:CASE_Nfin} the light curve of this 
system is phased separately with the pulsation period (0.08 d) and with the orbital period (2.83 d). 

Another two detached systems, N113 (OGLE-BLG-ECL-423112) and 
N121, are potentially interesting because of their CMD locations near the bottom of the 
red giant branch. If either of these turns out to be M22 member, it would provide a good 
reference point for isochrone fitting in $M-R$ and $M-L$ diagrams (see e.g. Kaluzny et al.
2013).  

\section{New data on known variables}
\label{sec:oldvar}

Of the 101 objects cataloged by C01-17 fourteen are located beyond our FOV, and two are pulsars 
without optical counterparts. Due to crowding and blending, among the eight variables 
discovered by PK03 in HST frames of the central part of M22 only PK-05 could have been identified (all 
star designations in this Section are taken from C01-17). For the remaining 78 stars 
membership status and membership probability were assigned using the criteria given in 
Section~\ref{sec:newmem}. Stars \#3, \#14, \#39, \#40, KT-01, KT-03, KT-05, KT-15, KT-18, 
KT-40, KT-41 and KT-48 turned out to be field objects. 

There are 10 RRab and 16 RRc pulsators in M22. A detailed analysis of our data on these 
objects will be published elsewhere; here we limit ourselves to a general remark
concerning the Blazhko effect. K13 suggest a  small incidence ($\sim$10\%) of the 
Blazhko effect  among RRab stars of M22, and do not detect any such effects in
stars of RRc type. In fact, the only star with a firmly established Blazhko effect they report 
is KT-55. We observe this behavior also in RRab stars \#2, \#3 and \#6. Another RRab star, 
\#23, suggested by K13 to have a rapidly changing or erratic period, does not show any such 
changes in our data: we only observe modest ($\pm$0.05 mag) variations of the descending 
shoulder of the light curve, which in principle might be interpreted as a weak Blazhko effect. 
Thus, according to our data, the incidence of the Blazhko effect among RRab stars is 40\% (50\% if 
\#23 is included). Moreover, we find a Blazhko effect of a varying strength in RRc stars \#18, 
\#19, \#25 and KT-36 (phase), \#15 (phase, shape) and KT-26, KT-37, Ku-1, Ku-2, Ku-3 and Ku-4 
(phase, shape, amplitude). Altogether, we observe Blazhko behavior for 15 (16) RR pulsators, 
i.e. an incidence rate of 58\% (62\%). Among the RRc stars the incidence is even higher -- 68\%. 
Thus, M22 is another GC with a large ($>$50\%) percentage of RRc Blazhko behavior, joining NGC 2808 
(Arellano Ferro et al. 2012) and M53 (Kunder et al. 2013b). 

\begin{figure}
   \centerline{\includegraphics[width=0.95\textwidth]{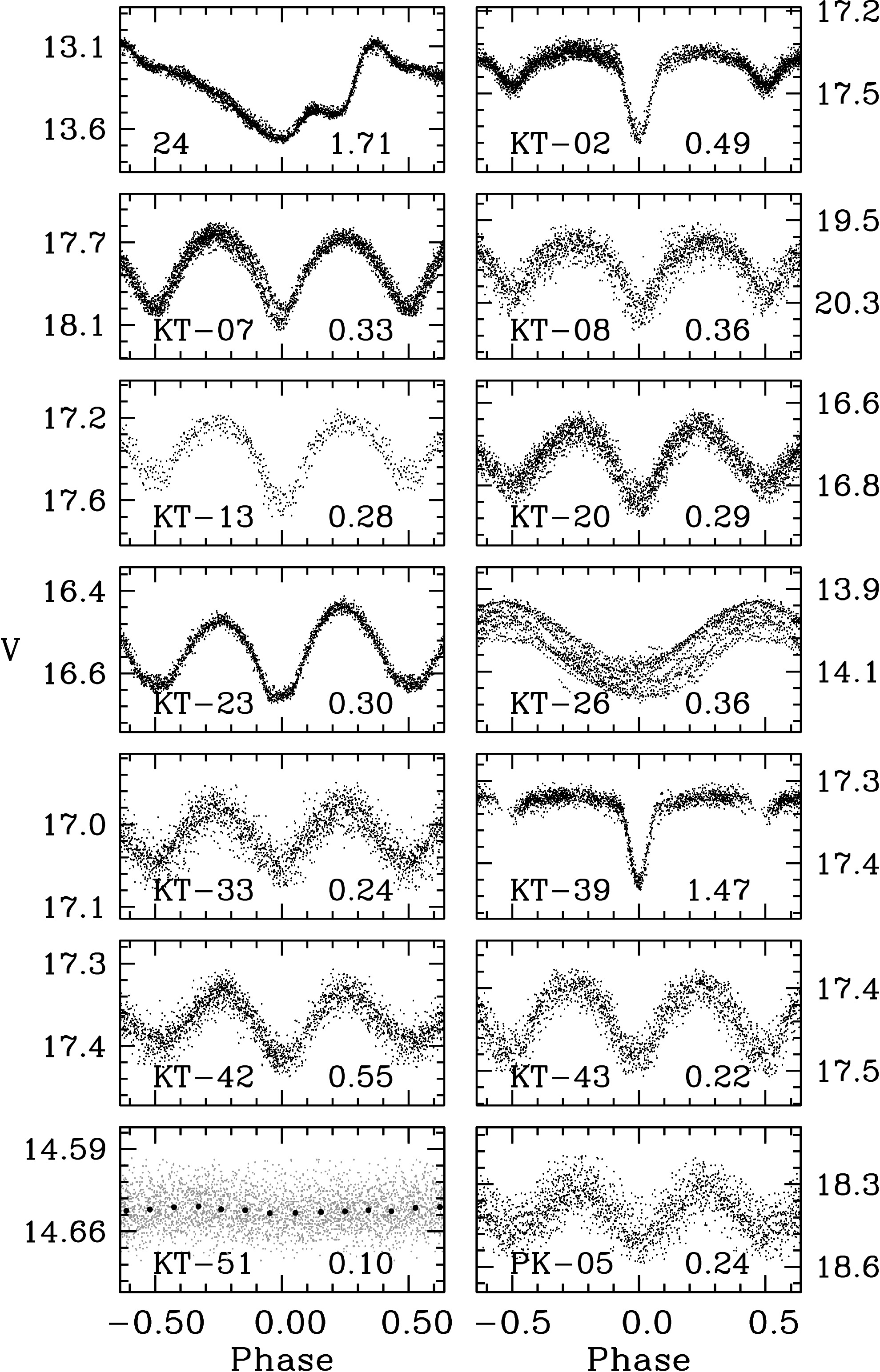}}
   \caption{Phased $V$-band light curves for a selection of the M22 variables 
    cataloged by C01-17. Phase-binned data are shown for KT-51 with heavy 
    black points. Individual panel labels give star ID from the C01-17 
    catalog, and period in days.
    \label{fig:Clement_fin}}
\end{figure}

Below we briefly describe C01-17 stars listed in Table \ref{tab:Clement}, whose light 
curves are shown in Fig. \ref{fig:Clement_fin}.

Star \#24: To our surprise, this object, listed as a non--variable by K13, turns out to be a
BL Her pulsator with $P = 1.715$ d, and a stable lightcurve. Our data show no period doubling 
phenomenon foreseen theoretically by Buchler \& Moskalik (1992), and for the first time 
observed by Smolec et al. (2012) in a star belonging to the Galactic bulge.

Star \#31: In our data no star closer than 5$''$ to the position of \#31 shows evidence for variability.   

KT-02: This Algol-type binary star, relatively isolated within M22, and located slightly above the turnoff of
the cluster, is a potentially valuable age and distance indicator (Kaluzny et al. 2005). KT01 
observed the $\sim$0.25 mag deep secondary minimum only. We find the primary minimum to be 
$\sim$0.4 mag deeper, indicating not too discrepant temperatures of the components. Thus, 
spectral lines of both the components should be visible, and despite the short period ($P=0.49$ d) 
the system is bright enough ($V=17.35$ mag) for good quality spectra to be obtained and an 
accurate velocity curve to be extracted. 

KT-26: The light curve of this star suggests that this is an RRc pulsator exhibiting the Blazhko 
effect. However, KT-26 is too blue to be an ordinary RRc star (both the $V$ and $B$ lightcurves 
are of very good quality, so that a large error in $B-V$ is rather unlikely, the more that our 
$V$-band brightness agrees very well with that of K13). The archival Hubble frame NGC6656-J9L948010 
reveals KT-26 is a $\sim$0$''$.75 blend of two stars with a flux ratio $\sim$15:8. Unfortunately, 
since this is the only available ACS frame taken in the F606W filter, one cannot tell which component of the 
blend is the proper variable. If the brighter one, then its brightness is lower by $\sim$0.5 mag  
than the combined brightness of the blend, and $\sim$0.2 mag lower than that of the weakest 
RR Lyr in M22 (i.e. star \#23). In that case, the proper variable would resemble the peculiar 
pulsator V37 in NGC 6362 (Smolec et al. 2017). If the fainter component were variable, 
the magnitude difference would increase to $\sim$1.2 mag and $\sim$0.9 mag, respectively,  
moving it to the BHB. Then, however, its period of 0.361366 d would be definitely too long for 
a BHB star. In any case, KT-26 clearly deserves  closer observational scrutiny.  

KT-39: Tentatively classified by KT01 as a contact binary, this is in fact another interesting 
and potentially valuable Algol-type system. Its location just below the subgiant branch indicates that at 
least one of the components must have left the main sequence, thus providing a good point 
for isochrone fitting. With a difference between depths of minima similar to that of KT-02, 
comparable isolation and a period three times longer, KT-39 is a relatively easy target for 
spectroscopy.   

KT-46: Another Algol-type binary. KT01 only observed the $\sim$1 mag deep primary minimum. We found the 
secondary minimum is more than ten times shallower, which together with a maximum brightness 
of $V\sim19.6$ mag and a period of only 0.61 d rather eliminates this system from the list of 
currently interesting objects. The light curve for KT-46 can be downloaded from the CASE archive.

KT-13, KT-20, KT-23, KT-33, KT-42 and KT-43 are contact binaries located at the turnoff or in 
the BS region (KT-42 was erroneously classified by KT01 as a possible pulsator). Another three 
contact binaries, KT-07, KT-08 and PK-05, occupy positions to the right of the lower main sequence.
For all the eight binaries complete light curves are presented. All of them, including PK-05 
which is placed closest to the center of M22, are well isolated within the cluster, so that 
radial velocity measurements seem entirely feasible (see Rozyczka et al. 2010). KT-08 is particularly interesting as the 
first, and so far the only, contact binary found within CASE to reside significantly ($\sim$2 mag) 
below the turnoff of a globular cluster.

KT-51: This star, located at the top of the EHB, was singled out by KT01
as the most interesting object in their sample; possibly a binary. We confirm its variability, 
however with a different period than theirs (0.103 d vs. $\sim$0.2 d) and with a different amplitude 
(0.04 mag vs. 0.06 mag). Thus, the question of the binarity of this object remains open. In the 
archival HST/WFPC2 frame UA2L0802M, KT-51 is an unresolved blend $\sim0''.3$ wide.

\Section{Summary}
 \label{sec:sum}
This contribution substantially increases the inventory of variable stars in the field 
of M22. A total of 359 variables or suspected variables were detected, 238 of which had been not 
known before. 102 members or likely PM--members of the cluster were identified, including  
20 SX~Phe, 10 RRab and 16 RRc pulsators, one BL Her pulsator, 21 contact binaries, and 8 detached 
or semi--detached binaries. Periods were obtained for almost all of the observed variables except 
a few cases with variability timescale longer than our time base. 

Among the new members of M22, the most interesting objects for follow--up studies are V125 -- a 
$\beta$~Lyr-type BHB binary, V129 -- a blue/yellow straggler with a W UMa-like light curve located 
halfway between EHB and RGB, and V134 -- an EHB object with $P=2.33$ d and sinusoidal light curve. 
Among nonmembers, observational scrutiny would be desirable for N107 -- a detached eclipsing 
binary hosting a $\delta$ Sct-type pulsator, N44 -- a contact binary whose luminosity seems to have 
decreased by 0.08 mag between 2000 and 2008, and N87 -- a peculiar $P=0.93$ d binary with ellipsoidal 
modulation and narrow minimum in the middle of one of the descending shoulders of the sinusoid 
which may be an optical counterpart of the Chandra X--ray source C183656.05-234845.5. Multimodality 
was detected in 16 SX Phe stars, with the blue straggler V112 being the most prominent example of 
this type of variability. 

We also provide substantial new data on the variables cataloged by C01-17. In particular, we identify 
M22 as the third GC with a large ($>$50\%) percentage of Blazhko effect incidence among RRc stars after 
NGC 2808 (Arellano Ferro et al. 2012) and M53 (Kunder et al. 2013b). The RRc star KT-26 shows a 
peculiar behaviour, resembling that found for V37 in NGC 6362 by Smolec et al. (2017). Finally, the
contact binary KT-08 is the first, and so far the only such system found within CASE to reside 
significantly ($\sim$2 mag) below the turnoff of a globular cluster. As such, it might provide 
some constraints on the evolution of binary systems in GCs. 

\Acknow
{We thank Grzegorz Pojma\'nski for the lc code which vastly facilitated the work with light curves, 
and to the anonymous referee for many comments and suggestions which substantially improved the 
manuscript. 
This paper is partly based on data obtained from the Mikulski Archive for Space Telescopes (MAST). 
STScI is operated by AURA, Inc., under NASA contract NAS5-26555. Support for MAST for non--HST data 
is provided by the NASA Office of Space Science via grant NNX09AF08G and by other grants and contracts.}

\section*{Appendix: V112 -- a multi-mode, non-radial SX Phe type pulsating star}

\subsection{Light curve decomposition}
The short period and unstable light curve of the blue straggler V112  suggests it to be
a SX Phe type star. Cores of globular clusters host many  such stars yet most appear 
to be of low  amplitude (Kaluzny 2000; for recent references see Nemec et al. 2017). 
Because of its large amplitude V112  seemed worthy of further attention. We performed a complete 
Fourier decomposition of its light curve, employing the NFIT code by one of authors (ASC). 
For early application of this code to SX Phe light curves, and underlying methods, see 
Mazur et al. (2003), and Schwarzenberg-Czerny (1999), respectively. The analysis is 
performed in stages, so that consecutive frequencies are identified in the periodogram, 
and subsequently data are prewhitened of them. In that way a Fourier model of the light 
curve is established. At the final stage the model is refined by fitting all frequency 
terms simultaneously by non--linear least squares, with adjustment of the base frequencies. 
The effective Nyquist interval of our observations is close to 130 c/d. Our decomposition 
of the light curve of V112  is {\em complete} in that we accounted for all frequencies in 
the range up to 100 c/d and with half--amplitudes over $0.0015$ mag, i.e. twice their 
typical standard deviation ($\sigma=0.008$ mag). Even for these small amplitudes the standard 
deviation of phases remains within $0.08P$ while for 9 strong modes they were $\ll 0.01P$. 

\subsection{Pulsation modes of V112 }
Our analysis revealed three base frequencies of pulsation, $f_0, f_1$, and $f_2$, with some 
harmonics and also seven combination frequencies between them (see Table \ref{t1}).
Hence it may be securely assumed all these frequencies correspond to the pulsation of V112.
The ratio $f_0/f_1=0.784$ is within the range of that for fundamental--to--first--overtone radial 
p--modes in SX Phe, depending on metallicity (e.g. Petersen \& Christensen--Dalsgaard 1996), 
hence it seems secure to identify $f_0$ and $f_1$ with the fundamental and first overtone 
pulsation of V112. If so, the presence of the combination mode $f_2+f_0$ with $f_2$ close 
to $f_0$ constitutes  evidence of a non--radial mode $f_2$.

In the light curve of V112, there appear another 3 seemingly unrelated frequencies $f_3, 
f_4$, and $f_5$. Note that $f_3$ differs substantially from the combination $2f_0-2f_2$ and its 
moon alias. It appears within a low frequency power bump at $f<2$ c/d. Such a bump does
appear in some SX Phe candidate stars observed by Kepler Satellite (Nemec et al. 2017), and a coherent frequency 
found in Kepler data for a $\delta$ Scuti star is interpreted as due to either stellar 
rotation or g--modes (Saio et al. 2015). Our ground--based mono--site data may suffer from a zero 
level drift for frequencies below 0.2 c/d (Mazur et al. 2003), though most of the 
low--frequency bump could be real, similar to the one in Kepler stars.
The remaining frequencies $f_4$ and $f_5$ are ill--expressed in our data. Although 
their amplitudes reach $6$ and $4.5\sigma$, an  NLSQ fit yields as standard deviation 
as large as that corresponding to a $0.1P$ uncertainity over a half time--span of our data. We leave 
the question of their reality and nature open.

Due to the simultaneous presence of well established fundamental and first overtone radial 
p--modes and a non--radial one, V112 belongs to a subgroup of SX Phe stars most suitable to 
an asteroseismic analysis. There are signs of the presence of low frequency $f_3 \& f_4$ 
oscillations consistent with g--modes, yet due to their small amplitude and uncertain fits
we refrain from further discussion. As $f_0$ and $f_0/f_1$ are tied to metallicity and 
luminosity, from such analysis of V112 it may be possible to get information on chemical 
composition and distance of M22 -- the more that the presence of non--radial mode(s) yields 
additional constraints.

\clearpage

\begin{table}[H]
\footnotesize
 \begin{center}
 \caption{\footnotesize Basic data of selected new variables discovered in the field of M22}
          \label{tab:CASE}
 \begin{tabular}{|l|c|c|c|c|c|c|c|c|}
  \hline
 ID & RA & DEC & $V$ & $B-V$ & $\Delta_V$ &Period & Type$^a$ & Mem$^c$\\
    & [deg] & [deg]  &[mag]& [mag] & [mag] & [d] &             & \\
  \hline
V112	&	279.10487	&	-23.90030	&	15.73	&	0.69	&	0.30	&	0.062316	&	SX	&	Y	\\
V116	&	279.14455	&	-23.96458	&	19.74	&	1.16	&	0.13	&	0.166970	&	$susp$	&	Y	\\
V117	&	279.06735	&	-23.93642	&	15.84	&	0.96	&	0.02	&	0.313255	&	$sin$	&	Y	\\
V125	&	278.98951	&	-23.79977	&	14.52	&	0.19	&	0.02	&	0.542888	&	EB	&	Y	\\
V129	&	279.08844	&	-23.86043	&	15.80	&	0.52	&	0.05	&	1.394800	&	EW	&	Y	\\
V130	&	279.07328	&	-23.95401	&	16.63	&	0.61	&	0.04	&	1.445972	&	EA	&	Y	\\
V131	&	279.10074	&	-23.90743	&	16.00	&	0.43	&	0.26	&	1.733622	&	EA	&	Y	\\
V133	&	279.15541	&	-23.89410	&	18.99	&	1.10	&	0.24	&	2.244228	&	EA	&	Y	\\
V134	&	279.06929	&	-23.93284	&	16.82	&	0.21	&	0.14	&	2.330917	&	$sin$	&	Y	\\
V135	&	279.23132	&	-23.92565	&	18.79	&	0.79	&	0.16	&	4.927996	&	EA	&	Y	\\
U39	&	279.06515	&	-23.97477	&	16.52	&	0.72	&	0.20	&	0.626613	&	EA	&	U	\\
U44	&	278.97674	&	-24.06404	&	18.95	&	1.25	&	0.27	&	0.811103	&    RS CVn/E	&	U	\\
U50	&	279.10452	&	-23.94817	&	18.29	&	1.15	&	0.11	&	1.922292	&	EA	&	U	\\
U51	&	279.16651	&	-23.87156	&	19.67	&	0.89	&	0.44	&	2.548926	&	EA	&	U	\\
U53	&	278.98998	&	-23.72693	&	16.70	&	0.78	&	0.61	&	4.148438	&	EA	&	U	\\
U56	&	279.16653	&	-23.84856	&	17.09	&	1.08	&	0.13	&	4.572758	&	$per$	&	U	\\
U61	&	279.22293	&	-23.78627	&	18.91	&	1.07	&	0.19	&	13.81400	&	EA	&	U	\\
U62	&	278.99650	&	-23.91997	&	16.76	&	0.84	&	0.04	&	20.75220	&	EA	&	U	\\
N04	&	279.11355	&	-23.79272	&	16.45	&	0.60	&	0.06	&	0.042513	&	SX/DSCT	&	N	\\
N10	&	278.97078	&	-24.06584	&	15.35	&	0.52	&	0.05	&	0.089940	&	SX/DSCT	&	N	\\
N11	&	279.11653	&	-23.82832	&	16.45	&	0.52	&	0.30	&	0.097236	&	DSCT 	&	N	\\
N12	&	279.14421	&	-24.01212	&	13.71	&	0.61	&	0.07	&	0.146750	&	DSCT	&	N	\\
N15	&	279.20356	&	-24.03031	&	16.76	&	0.46	&	0.20	&	0.245619	&	DSCT&	N	\\
N44	&	278.99070	&	-23.75487	&	15.91	&	0.87	&	0.19	&	0.414372	&	EW	&	N	\\
N65	&	279.21231	&	-24.04217	&	17.68	&	0.72	&	0.44	&	0.592321	&	EW	&	N	\\
N87	&	279.23363	&	-23.81226	&	17.03	&	1.47	&	0.06	&	0.925052	&	?   	&	N	\\
N107	&	279.17496	&	-24.00387	&	16.82	&	0.51	&	0.09	&	0.078948	&	DSCT 	&	N	\\
N107	&	279.17496	&	-24.00387	&	16.82	&	0.51	&	0.17	&	2.831980	&	EA	&	N	\\
N113	&	279.04831	&	-23.80348	&	17.23	&	0.91	&	0.20	&	3.997504	&	EA	&	N	\\
N121	&	279.19535	&	-23.87185	&	16.73	&	0.87	&	0.25	&	6.068994	&	EA	&	N	\\
  \hline
 \end{tabular}
\end{center}
{\footnotesize
$^a$ EA - detached eclipsing binary, EB - type $\beta$Lyr eclipsing binary, EW - contact binary, 
RS CVn/E - RS CVn-type binary with eclipse, SX - type SX Phe pulsator, DS - type $\delta$ Sct pulsator, 
$sin$ - sinusoidal light curve of unknown origin, $per$ - periodic variable of unknown type, $susp$ - suspected variable.\\
$^c$ Membership status: Y - member or likely member, U - no data or data ambiguous, N - field object.}
\end{table}

\clearpage

\begin{table}[H]
\footnotesize
 \begin{center}
 \caption{\footnotesize Basic data of C01-17 variables for which important new 
          information is provided}
          \label{tab:Clement}
 \begin{tabular}{|l|c|c|c|c|c|c|c|c|}
  \hline
 ID$^a$ & RA & DEC & $V$ & $B-V$ & $\Delta_V$ &Period & Type$^b$ & Mem$^d$\\ 
    & [deg] & [deg]  &[mag]& [mag] & [mag]    & [d]     & Remarks$^c$ & \\
  \hline
24	&	279.09078	&	-23.90371	&	13.42	&	0.90	&	0.65	&	1.714854	&	BL Her	&	Y	\\
KT-02	&	279.17424	&	-23.93922	&	17.33	&	0.65	&	0.28	&	0.490629	&	EA/EB	&	Y	\\
KT-07	&	279.15353	&	-23.95021	&	17.67	&	0.85	&	0.49	&	0.329797	&	EW	&	Y	\\
KT-08	&	279.14863	&	-23.92095	&	19.68	&	1.08	&	0.64	&	0.363902	&	EW	&	Y	\\
KT-13	&	279.12860	&	-23.89616	&	17.22	&	0.68	&	0.45	&	0.281733	&	EW	&	Y	\\
KT-20	&	279.10869	&	-23.85749	&	16.66	&	0.44	&	0.17	&	0.288495	&	EW; BS	&	Y	\\
KT-23	&	279.09932	&	-23.85456	&	16.47	&	0.43	&	0.24	&	0.298523	&	EW; BS	&	Y	\\
KT-26	&	279.09650	&	-23.88985	&	14.04	&	0.28	&	0.21	&	0.361366	&	RRc; Bl	&	Y	\\
KT-33	&	279.07025	&	-23.89847	&	16.96	&	0.72	&	0.08	&	0.244137	&	EW	&	Y	\\
KT-39	&	279.04161	&	-23.86605	&	17.28	&	0.84	&	0.19	&	1.474834	&	EA	&	Y	\\
KT-42	&	279.14445	&	-23.87534	&	17.29	&	0.62	&	0.11	&	0.554893	&	EW; BS	&	Y	\\
KT-43	&	279.10116	&	-23.93863	&	17.36	&	0.77	&	0.10	&	0.220520	&	EW	&	Y	\\
KT-46	&	279.09066	&	-23.97364	&	19.41	&	0.92	&	0.99	&	0.610198	&	EA	&	Y	\\
KT-51	&	279.14617	&	-23.88433	&	14.62	&	0.20	&	0.015	&	0.103422	&	$sin$; BHB	&	Y	\\
PK-05	&	279.09275	&	-23.90914	&	18.31	&	0.92	&	0.25	&	0.242839	&	EW	&	Y	\\
  \hline
 \end{tabular}
\end{center}
{\footnotesize 
$^a$ After C01-17.\\
$^b$ EA - detached eclipsing binary, EB - type $\beta$Lyr eclipsing binary, EW - contact binary,  
     BL Her - type BL Her pulsator, RRc - type RRc pulsator, sin - sinusoidal light curve of unknown origin.\\
$^c$ Bl - Blazhko effect, BHB - blue horizontal branch object, BS- blue straggler. \\
$^d$ Membership status: Y - member or likely member.}
\end{table}

\begin{table}[H]
\footnotesize
\begin{center}
\caption{Pulsation frequencies in V112  }\label{t1}
 \begin{tabular}[]{|r|c|c|}
\hline
Frequency[c/d] & Amplitude$^{a)}$ & Type \\
\hline
16.0472041~ & 0.1195 & $f_0(F)$ \\
32.0944082~ & 0.0440 & $2f_0$ \\
20.4731506~ & 0.0285 & $f_1(1~OT)$ \\
4.4259466~ & 0.0190 & $f_1-f_0$ \\
36.5203547~ & 0.0185 & $f_1+f_0$ \\
48.1416123~ & 0.0180 & $3f_0$ \\
11.6212575~ & 0.0135 & $2f_0-f_1$ \\
64.1888164~ & 0.0102 & $4f_0$ \\
52.5675588~ & 0.0098 & $2f_0+f_1$ \\
1.2720596: & 0.0076: & $f_3$ \\
68.6147629~ & 0.0070 & $3f_0+f_1$ \\
15.3964482~ & 0.0069 & $f_2 (NR)$ \\
27.6684616~ & 0.0052 & $3f_0-f_1$ \\
31.4436523~ & 0.0049 & $f_0+f_2$ \\
6.0188051: & 0.0048: & $f_4$ \\
84.6619670~ & 0.0043 & $4f_0+f_1$ \\
80.2360205~ & 0.0039 & $5f_0$ \\
91.9997502: & 0.0037: & $f_5$ \\
96.2832246~ & 0.0027 & $6f_0$ \\
112.3304287~ & 0.0020 & $7f_0$ \\
40.9463013~ & 0.0017 & $2f_1$ \\
61.4194519~ & 0.0015 & $3f_1$ \\
81.8926026~ & 0.0014 & $4f_1$ \\
128.3776328~ & 0.0014 & $8f_0$\\
\hline
\end{tabular}
\end{center}
\centering $^a$ Half peak-to-peak amplitude in magnitudes
\end{table}
\clearpage

\begin{figure}
    \centerline{\includegraphics[width=0.95\textwidth]{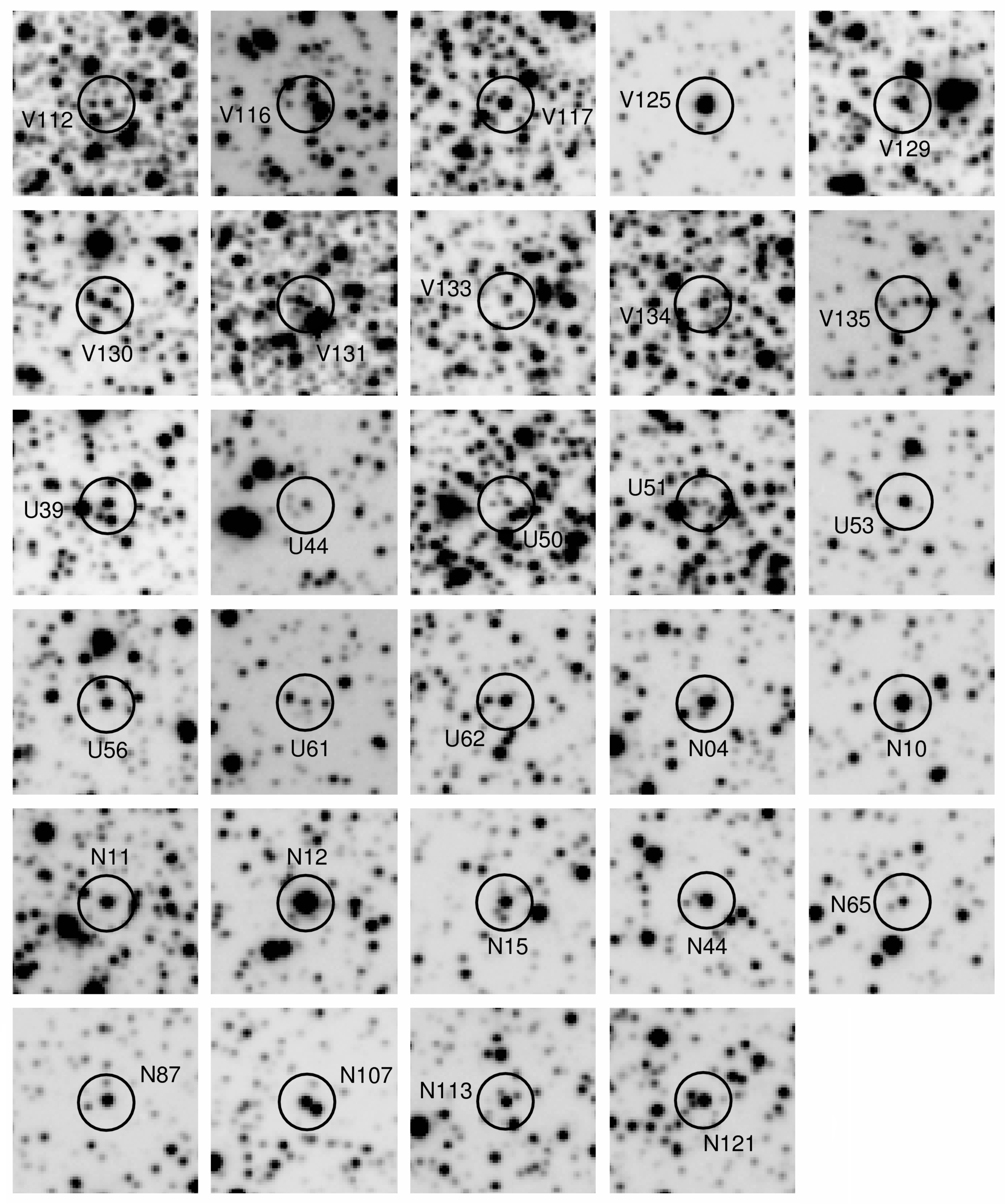}}
    \caption{Finding charts for the new variables whose light curves are shown
     in Figs. \ref{fig:CASE_Yfin}, \ref{fig:CASE_Ufin}, and \ref{fig:CASE_Nfin}. 
     Each chart is 30$''$ on a side. North is up and East to the left.
     \label{fig:maps}}
\end{figure}

\end{document}